\documentclass[10pt,conference]{IEEEtran}
\usepackage{cite}
\usepackage{amsmath}
\usepackage{amssymb}
\usepackage{amsfonts}
\usepackage{algorithmic}
\usepackage{graphicx}
\usepackage{textcomp}
\usepackage[dvipsnames]{xcolor}

\usepackage{hyperref}  
\hypersetup{pdfborder={0 0 0},colorlinks=true,urlcolor=blue,linkcolor=blue,citecolor=blue,bookmarks=true}

\hypersetup{pdftitle={Armadillo: An Efficient Framework for Numerical Linear Algebra}}
\hypersetup{pdfauthor={Conrad Sanderson and Ryan Curtin}}

\usepackage[utf8]{inputenc}
\usepackage[british]{babel}
\usepackage{hyphenat}
\usepackage[labelfont={small,bf},textfont=small,figurename={Fig.}]{caption}  

\usepackage{enumerate}
\usepackage{mathtools}
\usepackage{xspace}
\usepackage{booktabs}
\usepackage{algorithmic}
\usepackage{algorithm}
\usepackage{xfrac}
\usepackage{fancyvrb}  
\usepackage{adjustbox}

\usepackage[hang,flushmargin]{footmisc}   

\makeatletter
\def\footnoterule{\kern-3\p@
  \hrule \@width \columnwidth \kern 2.6\p@} 
\makeatother

\usepackage{newtxtext}  
\usepackage{newtxmath}  
\usepackage{inconsolata}  

\usepackage[absolute]{textpos}

\graphicspath{{./}{./figures/}}

\def\Vec#1{{\boldsymbol{#1}}}
\def\Mat#1{{\boldsymbol{#1}}}

\begin{document}

\title{\scalebox{0.79}{Armadillo: An Efficient Framework for Numerical Linear Algebra}}

\author
  {
  Conrad Sanderson~{$^{\dagger\ddagger}$} and Ryan Curtin~{$^{\diamond}$}\\
  ~\\
  {\normalsize{$^\dagger$} Data61~/~CSIRO, Australia}; {\normalsize{$^\ddagger$} Griffith University, Australia}; {\normalsize{$^\diamond$} NumFOCUS Inc., USA}
  }

\maketitle

\begin{abstract}

A major challenge in the deployment of scientific software solutions is
the adaptation of research prototypes to production-grade code.
While high-level languages like MATLAB are useful for rapid prototyping,
they lack the resource efficiency required for scalable production applications,
necessitating translation into lower level languages like C++.
Further, for machine learning and signal processing applications,
the underlying linear algebra primitives,
generally provided by the standard BLAS and LAPACK libraries,
are unwieldy and difficult to use,
requiring manual memory management and other tedium.
To address this challenge,
the Armadillo C++ linear algebra library provides an intuitive interface for writing linear algebra expressions
that are easily compiled into efficient production-grade implementations.
We describe the expression optimisations we have implemented in Armadillo,
exploiting template metaprogramming.
We demonstrate that these optimisations result in considerable efficiency gains
on a variety of benchmark linear algebra expressions.

\end{abstract}

\vspace{1ex}


\begin{textblock}{13.44}(1.28,14.80)
\hrule
\vspace{1ex}
\noindent
\scalebox{0.78}{\textbf{{$^\ast$}~Published in:} International Conference on Computer and Automation Engineering, pp.~303--307, 2025. ~~~ DOI:~\href{https://doi.org/10.1109/ICCAE64891.2025.10980539}{\tt 10.1109/ICCAE64891.2025.10980539}}
\end{textblock}

\vspace{-0.5ex}
\section{Introduction}
\vspace{0.5ex}

Deployment and productisation of various machine learning and signal processing algorithms
often requires conversion of research code written in a high-level language (eg.,~Matlab~\cite{Higham_2017})
into a lower level language such as {\small C} or {\small C++},
which is considerably more resource efficient~\cite{Stroustrup_2024}.
Resource efficiency is an important concern:
in datacenter environments, the efficiency of production code is
directly connected to cost (power costs and/or cloud resource costs).
In environments with constrained computational resources,
such as robots, unmanned aerial vehicles and spacecraft,
efficiency is especially important as prototype code may be entirely unable to run on the target device
due to limited memory or computational power.

Many algorithms inherently rely on numerical linear algebra operations,
which are typically provided by the well-tested industry standard {\small BLAS} and {\small LAPACK} toolkits~\cite{anderson1999lapack,Blackford_2002},
and their high-performance drop-in substitutes like {\small OpenBLAS}~\cite{OpenBLAS}, {\small MKL}~\cite{IntelMKL} and {\small AOCL}~\cite{AMD_AOCL}.
However, converting arbitrary linear~algebra expressions
into an efficient sequence of well-matched calls to {\small BLAS} and {\small LAPACK} routines is non-trivial~\mbox{\cite{Berenyi_2018,Psarras_2022}};
manual conversion can be laborious and error-prone,
and requires good understanding of the intricacies of {\small BLAS} and {\small LAPACK},
including various trade-offs across available routines and storage formats.
A further downside of directly using {\small BLAS}/{\small LAPACK} routines is that the resultant source code
is quite verbose,
has little similarity to the original mathematical expressions,
involves keeping track of many supporting variables,
and requires manual memory management.
Such aspects significantly reduce the readability of the source code,
raise the risk of bugs, and increase the maintenance burden~\cite{Sneed_2004,Malhotra_2016}.

To address the above issues in a coherent framework,
we have implemented the Armadillo linear algebra library for {\small C++}~\cite{Armadillo_JOSS_2016},
which automatically optimises mathematical expressions
(both at compile-time and run-time)
and efficiently maps them to {\small BLAS}/{\small LAPACK} routines,
all while providing a user-friendly Matlab-like programming interface directly in {\small C++}.
Armadillo essentially acts as a high-level {\it domain specific language}~\cite{Mernik_2005} built on top of the host {\small C++} language,
allowing for resource efficient numerical linear algebra without the many pain points of low-level code.
This enables rapid and low risk conversion of research code into production environments,
and even permits direct prototyping of algorithms within~{\small C++}.

As an expository demonstration of the reduced maintenance burden when using Armadillo,
consider the matrix expression $\Vec{c} = \Mat{A}^{-1}\Vec{b}$ 
for matrix $\Mat{A}$ and vectors $\Vec{b}$ and $\Vec{c}$,
which represents the solution to a system of linear equations.
Using Armadillo, it can be implemented directly in {\small C++} 
as a single readable and maintainable line of code:
{\small\tt vec c = inv(A) * b}.
Naively mapped, the above code will result in subsequent calls to three {\small LAPACK} and
{\small BLAS} functions%
\footnote
  {
  The first letter of each BLAS/LAPACK function is replaced with {\tt x} to express a set of functions
  that differ only in the associated element type (\textit{float}, \textit{double}, ...).
  For example, {\tt xGEMM} represents the {\tt SGEMM}, {\tt DGEMM}, {\tt CGEMM} and {\tt ZGEMM} functions.
  }%
: {\small \tt xGETRF}, {\small \tt xGETRI}, and {\small \tt xGEMV}.
Each of those three functions has between 6 and 11 parameters
and may require manual allocation of workspace memory.
In addition to hiding the verbosity and associated burdens with calls to {\small BLAS} and {\small LAPACK} functions,
Armadillo is also able to reinterpret the expression
and perform a better mapping to more efficient {\small BLAS}/{\small LAPACK} functions,
avoiding the explicit matrix inverse operation.

Armadillo employs two strategies for automatically optimising mathematical expressions,
both aiming to reduce computational effort:
\textbf{(i)} compile-time fusion of operations to reduce the need for temporary objects,
\textbf{(ii)} mixture of compile-time detection of expressions and run-time analysis of matrix properties,
with the aim of re-ordering and translating operations.
Both strategies extensively use {\small C++} template metaprogramming concepts~\cite{Czarnecki_2000,Vandevoorde_2017},
where the compiler is induced to reason at compile-time to generate code tailored for each expression.

We continue the paper as follows.
Section~\ref{sec:expr_opt} overviews the techniques for compile-time and run-time expression optimisation.
Section~\ref{sec:experiments} provides an empirical evaluation demonstrating the speedups obtained from the optimisations.
The salient points and avenues for further exploitation are summarised in Section~\ref{sec:conclusion}.

\newpage

\section{Expression Optimisation via Metaprogramming}
\label{sec:expr_opt}

Template metaprogramming induces the {\small C++} compiler
to run special programs written in a subset of the {\small C++} language.
Such metaprograms are executed entirely at compile-time,
and can be used to produce compiled code
that is specialised for arbitrary object and element types~\cite{Vandevoorde_2017}.

Rather than directly and immediately evaluating each component of a mathematical expression,
Armadillo exploits template metaprogramming via lightweight marker objects
that hold references to matrices and data associated with specific operations.
The marker objects are generated via user-accessible functions
(such as addition and multiplication)
and store the identifier of each operation as a custom \textit{type}
only visible to the {\small C++} compiler, rather than an explicit value.
The marker objects can be chained together,
leading to the full description of an arbitrary mathematical expression
to be visible to the {\small C++} compiler as an elaborate type,
comprised as a tree of operation types.
The evaluation of the entire expression is automatically performed
when it is assigned to a target matrix.
This approach is known as \textit{delayed evaluation}
(also known as \textit{lazy evaluation}),
and is in contrast to the traditional \textit{eager evaluation} and \textit{greedy evaluation} approaches~\cite{Watt_2004}.

As an illustrative example, let us consider the expression
{\tt
\begin{center}
Z = 0.4\hspace{0.4ex}{*}\hspace{0.4ex}X + 0.6\hspace{0.4ex}{*}\hspace{0.4ex}Y
\end{center}
}

\noindent
where {\small\tt X}, {\small\tt Y} and {\small\tt Z} are pre-defined {\small\tt Mat} objects,
each holding a $100{\times}100$ matrix.
In a traditional eager evaluation approach, 
the {\small\tt 0.4\hspace{0.4ex}{*}\hspace{0.4ex}X} operation would be evaluated first,
storing the intermediate result in a temporary matrix~{\small\tt T1}.
The {\small\tt 0.6\hspace{0.4ex}{*}\hspace{0.4ex}Y} operation would then result
in a secondary temporary matrix~{\small\tt T2}.
The temporary matrices {\small\tt T1} and {\small\tt T2} would then be added,
finally storing the result in matrix~{\small\tt Z}.
This approach for the evaluation of the entire expression is suboptimal and inefficient,
as it requires time-consuming memory allocation for the two temporary matrices
and three separate loops over the associated matrix elements.

The delayed evaluation approach implemented in Armadillo aims to address such inefficiencies.
Through overloading the~{\tt *}~operator function,
the operation {\small\tt 0.4\hspace{0.4ex}{*}\hspace{0.4ex}X} is not evaluated directly,
but is instead automatically converted to a lightweight templated marker object
named \mbox{\small\tt Op<Mat, op\_mul>},
which holds a reference to the~{\small\tt X} object
and a copy of the {\small\tt 0.4} scalar multiplier.
The nomenclature {\small\tt Op<...>} indicates that {\small\tt Op} is a {\small C++} template class,
with the items (types) between `{\small\tt <}' and `{\small\tt >}' specifying template parameters.
A similar {\small\tt Op} marker object is automatically constructed for the {\small\tt 0.6\hspace{0.4ex}{*}\hspace{0.4ex}Y} operation.
The~{\tt +}~operator function is overloaded to accept {\small\tt Mat} objects and arbitrary marker objects,
generating a templated {\small\tt Glue} marker object that holds references to the given objects.
In this example, it chains the two generated {\small\tt Op} objects,
resulting in the {\small\tt Glue} object having the following type:

\begin{center}
\scalebox{0.94}{\tt Glue< Op<Mat, op\_mul>, Op<Mat, op\_mul>, glue\_plus >}
\end{center}

The expression evaluation mechanism in Armadillo is then automatically invoked
through the {\small\tt =} operator defined in the {\small\tt Mat} object.
The mechanism interprets (at compile-time)
the nested types in the template parameters of the given {\small\tt Glue} object
and automatically generates compiled instructions equivalent to:

\begin{center}
\scalebox{0.90}{\tt for(int i=0; i<N; ++i) \{ Z[i] = 0.4\hspace{0.4ex}*\hspace{0.4ex}X[i] + 0.6\hspace{0.4ex}*\hspace{0.4ex}Y[i]; \}}
\end{center}

\noindent
where {\small\tt N} is the number of elements in matrices {\small\tt X}, {\small\tt Y} and~{\small\tt Z},
with {\small\tt X[i]} indicating the i-th element in matrix {\small\tt X}.
Apart from the lightweight {\small\tt Op} and {\small\tt Glue} marker objects
(which are automatically generated and pre-allocated at compile-time),
no other temporary objects are generated.
Furthermore, only one loop over the elements is required,
instead of three separate loops in the traditional eager evaluation approach.

As a further efficiency enhancement,
modern {\small C++} compilers exploit aggressive optimisation strategies
that are able to remove lightweight scaffolding objects.
This results in the compiler producing machine code
where the temporary {\small\tt Op} and {\small\tt Glue} objects are optimised away,
leaving only code absolutely necessary for the specialised loop, 
tailored for the given expression.
Moreover, this loop can be automatically \textit{vectorised} by the {\small C++} compiler,
where low-level \scalebox{0.95}{Single-Instruction-Multiple-Data~(SIMD)} instructions are exploited to achieve higher throughput~\cite{Stock_2012}.

The expression evaluation mechanisms in Armadillo include safety checks,
to ensure that only compatible sizes can be used for each given operation.
For example, checking that two matrices to be added or multiplied have conforming dimensions.

For mathematical expressions involving element-wise operations that can be chained,
the evaluation mechanism is able to handle an arbitrary number of components (eg., matrices) within the given expressions.
Other expressions are handled through detecting specific template patterns,
possibly embedded within longer expressions.
For example, the expression {\small\tt inv(A)\hspace{0.4ex}{*}\hspace{0.4ex}b}
is translated to the following {\small\tt Glue} template type:

\vspace{-0.5ex}
\begin{center}
\scalebox{0.90}{\tt Glue< Op<Mat, op\_inv>, Vec,  glue\_times >}
\end{center}
\vspace{-0.5ex}

\noindent
The above pattern is detected at compile-time,
and is automatically translated as a call to the {\small\tt xGESV} function in {\small LAPACK},
which solves a system of linear equations without the matrix inverse.

In general, expressions with matrix multiplication are typically translated
as calls to the {\small\tt xGEMM} and {\small\tt xGEMV} functions in {\small BLAS},
which are in turn multi-threaded and hand optimised for specific {\small CPU} architectures
in high-performance implementations such as {\small OpenBLAS}~\cite{OpenBLAS}.

Expression patterns are not necessarily blindly mapped to loops or {\small BLAS/LAPACK} functions.
Specific patterns are further analysed at run-time,
by analysing the properties of the constituent matrices.
For example, run-time analysis is used for detecting that in the expression {\small $\Mat{A}{\cdot}\Mat{A}^T$},
the matrix multiplication involves the same matrix and results in a symmetric matrix.
Rather than mapping the expression to the {\small\tt xGEMM} function by default,
the more efficient {\small\tt xSYRK} function can be used,
which exploits the symmetry property.

Analysis of matrix properties is also exploited in the evaluation of matrix multiplication chains.
For example, in the expression {\small $\Mat{A}{\cdot}\Mat{B}{\cdot}\Mat{C}{\cdot}\Mat{D}$},
each of the possible matrix pairs is examined.
The pair which results in the smallest matrix is multiplied first,
thereby reducing computational effort in subsequent matrix multiplications.
As such, it is possible for the entire expression to be evaluated right-to-left
(while respecting general non-commutativity of matrix multiplication),
rather than the traditional left-to-right order.

\clearpage

\section{Empirical Evaluation}
\label{sec:experiments}

We evaluate the following representative set of expressions to demonstrate some of the optimisations
automatically attainable by the expression processing frameworks implemented%
\footnote
  {
  Documentation for all the functionality available in Armadillo
  is available at \href{https://arma.sourceforge.net/docs.html}{\tt https://arma.sourceforge.net/docs.html}
  }
in the Armadillo library.

\vspace{1ex}

{
\fontsize{9.4}{10.4}\selectfont   
\begin{enumerate}[(1).]
\itemsep=0.5ex

\item
$\Mat{C} = 0.4{\cdot}\Mat{A}{~+~}0.6{\cdot}\Mat{B}$;
this is an instance of a compound expression involving
element-wise addition of matrices and element-wise multiplication of matrices by scalars.
A naive implementation evaluates each component separately,
generating temporary matrices for $0.4{\cdot}\Mat{A}$ and $0.6{\cdot}\Mat{B}$,
followed by adding the temporary matrices.
An optimised implementation is able to bypass the generation of the temporaries,
combining scalar multiplication and element addition into one loop
that can exploit high-performance SIMD instructions present in modern CPUs~\cite{Stock_2012}.
SIMD instructions such as AVX-512 allow efficient processing of chunks of data in one hit
instead of individual elements~\cite{Cebrian_2020}.

\item
$\Mat{C} = \Mat{A}_{(:,1)}^{~}{~+~}\Mat{B}_{(2,:)}^{T}$;
this expression involves element-wise addition of submatrices
(accessing individual columns and rows)
in conjunction with matrix transpose.
The notation $\Mat{A}_{(:,1)}$ denotes the first column of $\Mat{A}$,
while 
$\Mat{B}_{(2,:)}$ denotes the second row of $\Mat{B}$.
A naive implementation explicitly extracts the column and row into temporary vectors,
followed by applying a transpose operation that generates a further temporary vector,
which is then used for element-wise addition.
An optimised implementation bypasses the generation of all temporary vectors
as well as the explicit transpose operation,
and instead accesses the matrix elements directly,
performing an implicit transpose where required.

\item
$\Mat{C} = \text{diagmat}(\Mat{A}){\cdot}\Mat{B}$;
this expression demonstrates matrix multiplication where one of the matrices is converted to a diagonal matrix.
The $\text{diagmat}(\Mat{A})$ function indicates that all elements not on the main diagonal of $\Mat{A}$
are assumed to be zero.
In a naive implementation, the $\text{diagmat}(\Mat{A})$ function extracts the diagonal from $\Mat{A}$,
and places it a temporary matrix.
The temporary matrix (which is assumed by default to be dense) is then multiplied with $\Mat{B}$
through a call to the standard {\tt xGEMM} function in BLAS.
An optimised implementation omits generating the temporary,
and instead performs a specialised matrix multiplication
which exploits sparsity by assuming that only the diagonal elements of $\Mat{A}$ are non-zero.

\item
$\Mat{C} = \text{diagmat}(\Mat{A}{\cdot}\Mat{B})$;
in this expression the result of matrix multiplication is converted into a diagonal matrix.
A naive implementation would blindly evaluate $\Mat{A}{\cdot}\Mat{B}$ via the {\tt xGEMM} function in BLAS and store the result in a temporary matrix,
followed by extracting the diagonal from the temporary and placing it in the final result matrix.
An optimised implementation is able to determine that only the diagonal elements of the matrix multiplication are required,
thereby omitting unnecessary computations and temporaries.

\item
$k = \text{trace}(\Mat{A}{\cdot}\Mat{B})$;
this expression is similar to the preceding $\text{diagmat}(\Mat{A}{\cdot}\Mat{B})$ expression,
with the main difference that the diagonal elements of $\Mat{A}{\cdot}\Mat{B}$ are summed into the scalar~$k$.
In a naive implementation full matrix multiplication is performed,
while an optimised implementation performs a partial matrix multiplication to obtain only the diagonal elements.

\item
$\Mat{E} = \Mat{A}_{m{\times}m} \cdot \Mat{B}_{m{\times}\frac{m}{2}} \cdot \Mat{C}_{\frac{m}{2}{\times}\frac{m}{3}} \cdot \Mat{D}_{\frac{m}{3}{\times}\frac{m}{4}}$.
this is an instance of chained matrix multiplication resulting in a matrix.
Here the matrices are progressively decreasing in size.
A~naive implementation would evaluate each of the matrix products in the standard left-to-right manner,
disregarding the wider context of the expression.
An optimised implementation can examine the sizes of all possible matrix products within the expression,
and determine that evaluating the products in a reversed order will save computational effort.

\item
$k = \Vec{a}^T \cdot \text{diagmat}(\Mat{B}) \cdot \Vec{c}$;
this is an example of chained matrix multiplication that results in a scalar value,
where $\Vec{a}$ and $\Vec{c}$ are column vectors.
A~naive implementation computes each component separately
(matrix transpose and generation of diagonal matrix)
resulting in temporary matrices,
and then performs matrix multiplication involving the temporaries.
An optimised implementation can examine the expression
and determine that only a single and straightforward element-wise multiply-and-sum loop is required
over the underlying components,
avoiding unnecessary computations and generation of temporaries.
This type of expression optimisation is invoked in Armadillo via the {\footnotesize\tt as\_scalar()} function.

\item
$\Mat{B} = \Mat{A} \cdot \Mat{A}^{T}$;
this expression is seemingly straightforward,
involving a matrix being multiplied with its transposed version,
resulting in a symmetric matrix.
A naive implementation disregards this fact
and blindly calculates the matrix product
by treating the two components as separate matrices after an explicit transpose operation.
A semi-optimised implementation can avoid the explicit transpose
by appropriate mapping to the {\tt xGEMM} function in BLAS.
However, a fully optimised implementation can detect that the two matrices to be multiplied are the same,
and map the expression to the more efficient {\tt xDSYRK} function in BLAS,
which exploits the symmetry aspect and avoids unnecessary computations.

\item
$\Mat{C} = \Mat{A}^{-1} \cdot \Vec{b}$;
this expression indicates that a solution to a system of linear equations is \textit{implicitly} sought.
A naive implementation ignores the intent of the expression and calculates the inverse of matrix $\Mat{A}$
followed by a matrix multiplication.
Calculating the inverse is not only computationally inefficient, but also potentially numerically unstable.
An optimised implementation can detect the intent of the expression and map it to the more appropriate {\tt xGESV} function in LAPACK,
which finds the solution through a more numerically stable algorithm~\cite{anderson1999lapack}.

\item
$\Mat{C} = \text{solve}(\Mat{A},\Vec{b})$ where $\Mat{A}$ is a tri-diagonal band matrix;
this expression indicates that a solution to a system of linear equations is \textit{explicitly} sought,
with $\Mat{A}$ having a special sparse structure.
A~naive implementation would disregard the structure.
An optimised implementation can analyse the matrix and choose a more tailored solver function in LAPACK,
thereby exploiting the sparse structure to avoid superfluous computations.

\end{enumerate}
}

\vspace{0.5ex}

{
\fontsize{10.1}{11.1}\selectfont

\noindent
For each of the above expressions,
the following multiple matrix sizes are used, ranging from small to large:
{\small $\{~100\times100$, $250\times250$, $500\times500$, $1000\times1000~\}$}.
The evaluation is done on a machine with an {\small AMD} Ryzen {\small 7640U} {\small x86-64} {\small CPU} running at {\small 3.5~GHz}.
All source code was compiled with the {\small GCC} {\small 14.2} {\small C++} compiler.
We also used the open-source {\small OpenBLAS} {\small 0.3.26} library
which provides optimised implementations of {\small BLAS} and {\small LAPACK} routines~\cite{OpenBLAS}.

The results shown in Fig.~\ref{fig:results} demonstrate that the optimised handling of expressions in Armadillo
leads to considerable reduction in computational effort.
Across the considered expressions, the reduction in wall-clock time is often over 50\%,
and in several cases it is over 90\%.

Fig.~\ref{fig:example_prog} shows a simple Armadillo-based C++ program to demonstrate its intuitive programming syntax.
Fig.~\ref{fig:trace} lists a trace of corresponding internal function calls,
hiding from the user the complexity of calling {\small BLAS} and {\small LAPACK} functions.

}

\begin{figure*}[t!]
\small
\begin{minipage}{1\textwidth}
\begin{minipage}{0.48\textwidth}
\centering

{(1) expression: $\Mat{C} = 0.4{\cdot}\Mat{A} + 0.6{\cdot}\Mat{B}$}
\vspace{1ex}

\begin{tabular}{cccc}
\hline
{\bf matrix size} & {\bf naive}            & {\bf optimised}           & {\bf reduction} \\ \hline
 100$\times$100   & $5.51 \times 10^{-6}$  & $   2.26 \times 10^{-6}$  & 59.04\%         \\
 250$\times$250   & $4.04 \times 10^{-5}$  & $   1.66 \times 10^{-5}$  & 58.89\%         \\
 500$\times$500   & $1.87 \times 10^{-4}$  & $   6.95 \times 10^{-5}$  & 62.87\%         \\
1000$\times$1000  & $2.45 \times 10^{-3}$  & $   7.85 \times 10^{-4}$  & 67.90\%         \\ \hline
\end{tabular}

\end{minipage}
\hfill
\begin{minipage}{0.48\textwidth}
\centering

{(6) expression: $\Mat{E} = \Mat{A}_{m{\times}m} \cdot \Mat{B}_{m{\times}\frac{m}{2}} \cdot \Mat{C}_{\frac{m}{2}{\times}\frac{m}{3}} \cdot \Mat{D}_{\frac{m}{3}{\times}\frac{m}{4}}$}
\vspace{0.5ex}

\begin{tabular}{cccc}
\hline
{\bf matrix size} & {\bf naive}            & {\bf optimised}        & {\bf reduction} \\ \hline
 100$\times$100   & $1.20 \times 10^{-5}$  & $6.17 \times 10^{-6}$  & 48.53\%         \\
 250$\times$250   & $2.05 \times 10^{-4}$  & $1.02 \times 10^{-4}$  & 50.20\%         \\
 500$\times$500   & $1.54 \times 10^{-3}$  & $7.94 \times 10^{-4}$  & 48.34\%         \\
1000$\times$1000  & $1.21 \times 10^{-2}$  & $6.02 \times 10^{-3}$  & 50.17\%         \\ \hline
\end{tabular}

\end{minipage}
\end{minipage}

\vspace{3ex}

\begin{minipage}{1\textwidth}
\begin{minipage}{0.48\textwidth}
\centering

{(2) expression: $\Mat{C} = \Mat{A}_{(:,1)}^{~} + \Mat{B}_{(2,:)}^{T}$}
\vspace{0.2ex}

\begin{tabular}{cccc}
\hline
{\bf matrix size} & {\bf naive}            & {\bf optimised}        & {\bf reduction} \\ \hline
 100$\times$100   & $9.52 \times 10^{-8}$  & $3.38 \times 10^{-8}$  & 64.50\%         \\
 250$\times$250   & $2.94 \times 10^{-7}$  & $1.05 \times 10^{-7}$  & 64.43\%         \\
 500$\times$500   & $7.01 \times 10^{-7}$  & $3.37 \times 10^{-7}$  & 51.94\%         \\
1000$\times$1000  & $1.30 \times 10^{-6}$  & $7.38 \times 10^{-7}$  & 43.07\%         \\ \hline
\end{tabular}

\end{minipage}
\hfill
\begin{minipage}{0.48\textwidth}
\centering

{(7) expression: $k = \Vec{a}^T \cdot \text{diagmat}(\Mat{B}) \cdot \Vec{c}$}
\vspace{1ex}

\begin{tabular}{cccc}
\hline
{\bf matrix size} & {\bf naive}            & {\bf optimised}         & {\bf reduction} \\ \hline
 100$\times$100   & $1.85 \times 10^{-6}$  & $7.21 \times 10^{-10}$  & 99.96\%         \\
 250$\times$250   & $1.14 \times 10^{-5}$  & $8.54 \times 10^{-10}$  & 99.99\%         \\
 500$\times$500   & $4.77 \times 10^{-5}$  & $7.21 \times 10^{-10}$  & 99.99\%         \\
1000$\times$1000  & $1.99 \times 10^{-4}$  & $7.24 \times 10^{-10}$  & 99.99\%         \\ \hline
\end{tabular}

\end{minipage}
\end{minipage}

\vspace{3ex}

\begin{minipage}{1\textwidth}
\begin{minipage}{0.48\textwidth}
\centering

{(3) expression: $\Mat{C} = \text{diagmat}(\Mat{A}) \cdot \Mat{B}$}
\vspace{1ex}

\begin{tabular}{cccc}
\hline
{\bf matrix size} & {\bf naive}            & {\bf optimised}        & {\bf reduction} \\ \hline
 100$\times$100   & $3.84 \times 10^{-5}$  & $2.80 \times 10^{-6}$  & 92.70\%         \\
 250$\times$250   & $6.49 \times 10^{-4}$  & $2.81 \times 10^{-5}$  & 95.67\%         \\
 500$\times$500   & $5.01 \times 10^{-3}$  & $2.01 \times 10^{-4}$  & 95.99\%         \\
1000$\times$1000  & $4.14 \times 10^{-2}$  & $1.49 \times 10^{-3}$  & 96.40\%         \\ \hline
\end{tabular}

\end{minipage}
\hfill
\begin{minipage}{0.48\textwidth}
\centering

{(8) expression: $\Mat{B} = \Mat{A} \cdot \Mat{A}^{T}$}
\vspace{1ex}

\begin{tabular}{cccc}
\hline
{\bf matrix size} & {\bf naive}            & {\bf optimised}        & {\bf reduction} \\ \hline
 100$\times$100   & $3.97 \times 10^{-5}$  & $3.35 \times 10^{-5}$  & 15.59\%         \\
 250$\times$250   & $6.65 \times 10^{-4}$  & $3.78 \times 10^{-4}$  & 43.19\%         \\
 500$\times$500   & $5.07 \times 10^{-3}$  & $2.67 \times 10^{-3}$  & 47.41\%         \\
1000$\times$1000  & $4.32 \times 10^{-2}$  & $2.21 \times 10^{-2}$  & 48.89\%         \\ \hline
\end{tabular}

\end{minipage}
\end{minipage}

\vspace{3ex}

\begin{minipage}{1\textwidth}
\begin{minipage}{0.48\textwidth}
\centering

{(4) expression: $\Mat{C} = \text{diagmat}(\Mat{A} \cdot \Mat{B})$}
\vspace{1ex}

\begin{tabular}{cccc}
\hline
{\bf matrix size} & {\bf naive}            & {\bf optimised}        & {\bf reduction} \\ \hline
 100$\times$100   & $3.88 \times 10^{-5}$  & $4.86 \times 10^{-6}$  & 87.47\%         \\
 250$\times$250   & $6.51 \times 10^{-4}$  & $3.99 \times 10^{-5}$  & 93.87\%         \\
 500$\times$500   & $5.02 \times 10^{-3}$  & $1.75 \times 10^{-4}$  & 96.51\%         \\
1000$\times$1000  & $4.11 \times 10^{-2}$  & $1.93 \times 10^{-3}$  & 95.31\%         \\ \hline
\end{tabular}

\end{minipage}
\hfill
\begin{minipage}{0.48\textwidth}
\centering

{(9) expression: $\Mat{C} = \Mat{A}^{-1} \cdot \Vec{b}$}
\vspace{1ex}

\begin{tabular}{cccc}
\hline
{\bf matrix size} & {\bf naive}            & {\bf optimised}        & {\bf reduction} \\ \hline
 100$\times$100   & $1.47 \times 10^{-4}$  & $5.45 \times 10^{-5}$  & 62.92\%         \\
 250$\times$250   & $1.46 \times 10^{-3}$  & $4.69 \times 10^{-4}$  & 67.91\%         \\
 500$\times$500   & $8.23 \times 10^{-3}$  & $2.79 \times 10^{-3}$  & 66.16\%         \\
1000$\times$1000  & $5.33 \times 10^{-2}$  & $1.90 \times 10^{-2}$  & 64.34\%         \\ \hline
\end{tabular}

\end{minipage}
\end{minipage}

\vspace{3ex}

\begin{minipage}{1\textwidth}
\begin{minipage}{0.48\textwidth}
\centering

{(5) expression: $k = \text{trace}(\Mat{A} \cdot \Mat{B})$}
\vspace{1ex}

\begin{tabular}{cccc}
\hline
{\bf matrix size} & {\bf naive}            & {\bf optimised}         & {\bf reduction} \\ \hline
 100$\times$100   & $3.73 \times 10^{-5}$  & $5.98 \times 10^{-11}$  & 99.99\%         \\
 250$\times$250   & $6.42 \times 10^{-4}$  & $6.62 \times 10^{-11}$  & 99.99\%         \\
 500$\times$500   & $4.93 \times 10^{-3}$  & $6.71 \times 10^{-11}$  & 99.99\%         \\
1000$\times$1000  & $4.03 \times 10^{-2}$  & $2.71 \times 10^{-10}$  & 99.99\%         \\ \hline
\end{tabular}

\end{minipage}
\hfill
\begin{minipage}{0.48\textwidth}
\centering

{(10) expression: $\Mat{C} = \text{solve}(\Mat{A},\Vec{b})$ where $\Mat{A}$ is a tri-diagonal}
\vspace{1ex}

\begin{tabular}{cccc}
\hline
{\bf matrix size} & {\bf naive}            & {\bf optimised}        & {\bf reduction} \\ \hline
 100$\times$100   & $8.11 \times 10^{-5}$  & $2.16 \times 10^{-5}$  & 73.40\%         \\
 250$\times$250   & $6.19 \times 10^{-4}$  & $7.40 \times 10^{-5}$  & 88.04\%         \\
 500$\times$500   & $3.31 \times 10^{-3}$  & $2.06 \times 10^{-4}$  & 93.77\%         \\
1000$\times$1000  & $2.13 \times 10^{-2}$  & $1.30 \times 10^{-3}$  & 93.91\%         \\ \hline
\end{tabular}

\end{minipage}
\end{minipage}

\caption
  {
  \small
  Comparison of time taken (in seconds) for various matrix expressions,
  using naive (non-optimised) and automatically optimised implementations
  within the Armadillo linear algebra library.
  Average wall-clock time across 1000 runs is reported.
  Evaluations were performed on an AMD Ryzen 7640U CPU, running at 3.5~GHz.
  Code was compiled with the GCC 14.2 C++ compiler with the following flags: \texttt{\small -O3~-march=native}.
  OpenBLAS 0.3.26 was used for optimised implementations of BLAS and LAPACK routines~\cite{OpenBLAS}.
  }
\label{fig:results}
\vspace{-2ex}
\end{figure*}

\begin{figure*}[t!]
\begin{minipage}{1\textwidth}
\begin{minipage}{0.485\textwidth}
\footnotesize
\hrule
\vspace{1ex}
\begin{verbatim}
01: #include <armadillo>
02: 
03: using namespace arma;
04: 
05: int main()
06:   {
07:   // generate random 100x100 matrix
08:   mat A(100, 100, fill::randu);
09:   
10:   // generate random 100x1 vector
11:   vec b(100, fill::randu);
12:   
13:   // solve for x in random symmetric system AA'x = b
14:   vec x = solve( A * A.t(), b );
15:   
16:   x.print("x:");
17:   
18:   return 0;
19:   }
\end{verbatim}
\hrule
\vspace{0.5ex}
\caption{A simple Armadillo-based C++ program, solving a random symmetric system of linear equations.}
\vspace{0.8ex}
\label{fig:example_prog}
\end{minipage}
\hfill
\begin{minipage}{0.485\textwidth}
\tiny
\hrule
\vspace{0.5ex}
\begin{verbatim}
Op<T1, op_type>::Op(T1&) [T1 = Mat; op_type = op_htrans]
operator*(T1&, T2&) [T1 = Mat; T2 = Op<Mat,op_htrans>]
Glue<T1, T2, glue_type>::Glue(T1&, T2&) [T1 = Mat; T2 = Op<Mat,op_htrans>; glue_type = glue_times]
solve(Base<double, T1>&, Base<double, T2>&)
Glue<T1, T2, glue_type>::Glue(T1&, T2&) [... glue_type = glue_solve_gen_def]
Col::Col(Base<double,T1>&) [T1 = Glue<Glue<Mat,Op<Mat,op_htrans>,glue_times>,Mat,glue_solve_gen_def>]
Mat::operator=(Glue<T1, T2, glue_type>&) [... glue_type = glue_solve_gen_def]
glue_solve_gen_def::apply(Mat&, Glue<T1, T2, glue_solve_gen_def>&)
glue_solve_gen_full::apply(Mat&, Base<double, T1>&, Base<double, T2>&, uword)
Mat::Mat(Glue<T1, T2, glue_type>&) [T1 = Mat; T2 = Op<Mat,op_htrans>; glue_type = glue_times]
glue_times::apply(Mat&, Glue<T1, T2, glue_times>&) [T1 = Mat; T2 = Op<Mat,op_htrans>]
glue_times_redirect<2>::apply(Mat&, Glue<T1, T2, glue_times>&) [T1 = Mat; T2 = Op<Mat,op_htrans>]
glue_times::apply(Mat&, TA&, TB&, double) [trans_A = false; trans_B = true; TA = Mat; TB = Mat]
Mat::set_size(uword, uword) [uword = long long unsigned int] [in_n_rows: 100; in_n_cols: 100]
Mat::init(): acquiring memory
blas::syrk(...)
glue_solve_gen_full::apply(): detected square system
band_helper::is_band(uword&, uword&, Mat&, uword) [uword = long long unsigned int]
trimat_helper::is_triu(Mat&)
trimat_helper::is_tril(Mat&)
glue_solve_gen_full::apply(): rcond + sym
auxlib::solve_sym_rcond(Mat&, double&, Mat&, Base<double, T1>&) [T1 = Mat; ...]
Mat::operator=(Mat&) [this: e0a67920; in_mat: e0a67860]
Mat::init_warm(uword, uword) [uword = long long unsigned int] [in_n_rows: 100; in_n_cols: 1]
Mat::init(): acquiring memory
lapack::lansy(...)
lapack::sytrf(...)
lapack::sytrs(...)
lapack::sycon(...)
Mat::destructor: releasing memory
\end{verbatim}
\vspace{-2ex}
\hrule
\vspace{0.5ex}
\caption
  {
  \small
  An abridged trace of internal function calls and \mbox{debugging} messages
  resulting from line 14 in Fig.~\ref{fig:example_prog},
  containing the expression \mbox{\footnotesize\tt vec x = solve( A * A.t(), b )}.
  }
\label{fig:trace}
\end{minipage}
\end{minipage}
\end{figure*}

\clearpage
\clearpage

\section{Conclusion}
\label{sec:conclusion}

Armadillo facilitates easy and maintainable representation of arbitrary linear algebra expressions in C++
that are efficiently mapped to underlying BLAS and LAPACK operations.
Users do not need to worry about cumbersome manual memory management
or complicated calls to BLAS and LAPACK subroutines.
There is virtually no performance penalty for the abstractions provided by Armadillo.
Moreover, through under-the-hood template metaprogramming and \mbox{automatic} optimisations of expressions,
Armadillo can achieve considerable reductions in processing time over direct and/or naive implementations.

Work on Armadillo started in 2008.
Over the years the library has been iteratively and collaboratively developed
with feedback from the wider scientific and engineering communities.
The library provides over 200 functions;
in addition to elementary operations,
there are functions for statistics, signal processing, non-contiguous submatrix views, and various matrix factorisations.
The library is currently comprised of about 135,000 lines of templated code,
excluding BLAS and LAPACK routines.
Support is provided for matrices with single- and double-precision floating point elements (in both real and complex forms),
as well as integer elements.
Dense and sparse storage formats are supported.

Armadillo is now in a mature state and in wide production use.
For example, Armadillo has been successfully used to accelerate computations in open-source projects
such as
the {\it ensmallen} library for numerical optimisation~\cite{ensmallen2021}
and
the {\it mlpack} library for machine learning~\cite{mlpack2023},
which provide production-ready applications
for a variety of environments,
including low-resource devices such as small microcontrollers.
Armadillo has also been used for accelerating 
over 1000 packages for the R statistical environment~\cite{rcpparma}.

Armadillo can be obtained from \scalebox{0.8}{\href{https://arma.sourceforge.net}{\tt https://arma.sourceforge.net}},
with the source code provided under the permissive Apache~2.0 license~\cite{Laurent_2004,Li_2025},
which allows unencumbered use in commercial products.
Armadillo is also included as part of all major Linux distributions.

In future work we plan to extend Armadillo to include support 
for half-precision floating point (FP16)
and `brain floating point' (BF16) element types~\cite{Henry_2019},
as well as to bring the same kinds of expression optimisations to
GPU-based linear algebra via the companion Bandicoot library~\cite{Curtin_2025}.

\vspace{3ex}

\section*{Acknowledgements}

{
\small
The authors would like to thank the wider open-source community
as well as all bug reporters and contributors to Armadillo;
without them this work would not have been possible.
To explicitly promote correct and safe functionality,
the Armadillo codebase was developed entirely without
the use of Large Language Models (LLMs)~\cite{Hicks_2024}.
Ryan~Curtin's contributions are based on work supported
by the National Aeronautics and Space Administration (NASA)
under the ROSES-23 HPOSS program, grant no.~80NSSC24K1524.
}

\newpage
\bibliographystyle{IEEEtran_mod}
\bibliography{refs}

\begin{thebibliography}{10}
\itemsep=0.65ex
\providecommand{\url}[1]{#1}
\csname url@samestyle\endcsname
\providecommand{\newblock}{\relax}
\providecommand{\bibinfo}[2]{#2}
\providecommand{\BIBentrySTDinterwordspacing}{\spaceskip=0pt\relax}
\providecommand{\BIBentryALTinterwordstretchfactor}{4}
\providecommand{\BIBentryALTinterwordspacing}{\spaceskip=\fontdimen2\font plus
\BIBentryALTinterwordstretchfactor\fontdimen3\font minus
  \fontdimen4\font\relax}
\providecommand{\BIBforeignlanguage}[2]{{%
\expandafter\ifx\csname l@#1\endcsname\relax
\typeout{** WARNING: IEEEtran.bst: No hyphenation pattern has been}%
\typeout{** loaded for the language `#1'. Using the pattern for}%
\typeout{** the default language instead.}%
\else
\language=\csname l@#1\endcsname
\fi
#2}}
\providecommand{\BIBdecl}{\relax}
\BIBdecl

\bibitem{Higham_2017}
D.~J. Higham and N.~J. Higham, \emph{{MATLAB} Guide}, 3rd~ed.\hskip 1em plus
  0.5em minus 0.4em\relax SIAM, 2017.

\bibitem{Stroustrup_2024}
B.~Stroustrup, \emph{Programming: Principles and Practice Using {C++}},
  3rd~ed.\hskip 1em plus 0.5em minus 0.4em\relax Addison-Wesley, 2024.

\bibitem{anderson1999lapack}
E.~Anderson, Z.~Bai, C.~Bischof, S.~Blackford, J.~Demmel \emph{et~al.},
  \emph{{LAPACK}~Users' Guide}.\hskip 1em plus 0.5em minus 0.4em\relax SIAM,
  1999.

\bibitem{Blackford_2002}
L.~S. Blackford, J.~Demmel, J.~Dongarra, I.~Duff, S.~Hammarling \emph{et~al.},
  ``An updated set of basic linear algebra subprograms ({BLAS}),'' \emph{ACM
  Trans.~Mathematical Software}, vol.~28, no.~2, pp. 135--151, 2002.

\bibitem{OpenBLAS}
\BIBentryALTinterwordspacing
Z.~Xianyi and M.~Kroeker, ``{OpenBLAS}: An optimized {BLAS} library,'' 2025.
  [Online]. Available: \url{http://www.openmathlib.org/OpenBLAS/}
\BIBentrySTDinterwordspacing

\bibitem{IntelMKL}
\BIBentryALTinterwordspacing
{Intel Corporation}, ``{Math Kernel Library (MKL)},'' 2025. [Online].
  Available: \url{https://software.intel.com/mkl}
\BIBentrySTDinterwordspacing

\bibitem{AMD_AOCL}
\BIBentryALTinterwordspacing
{Advanced Micro Devices}, ``{AMD Optimizing CPU Libraries~(AOCL)},'' 2024.
  [Online]. Available: \url{https://www.amd.com/en/developer/aocl.html}
\BIBentrySTDinterwordspacing

\bibitem{Berenyi_2018}
D.~Ber\'{e}nyi, A.~Leitereg, and G.~Lehel, ``Towards scalable pattern-based
  optimization for dense linear algebra,'' \emph{Concurrency and Computation:
  Practice and Experience}, vol.~30, no.~22, 2018.

\bibitem{Psarras_2022}
C.~Psarras, H.~Barthels, and P.~Bientinesi, ``The linear algebra mapping
  problem: Current state of linear algebra languages and libraries,'' \emph{ACM
  Transactions on Mathematical Software}, vol.~48, no.~3, 2022.

\bibitem{Sneed_2004}
H.~M. Sneed, ``A cost model for software maintenance \& evolution,'' in
  \emph{IEEE International Conference on Software Maintenance}, 2004.

\bibitem{Malhotra_2016}
R.~Malhotra and A.~Chug, ``Software maintainability: Systematic literature
  review and current trends,'' \emph{International Journal of Software
  Engineering and Knowledge Engineering}, vol.~26, no.~8, pp. 1221--1253, 2016.

\bibitem{Armadillo_JOSS_2016}
C.~Sanderson and R.~Curtin, ``Armadillo: a template-based {C++} library for
  linear algebra,'' \emph{Journal of Open Source Software}, vol.~1, no.~2,
  p.~26, 2016.

\bibitem{Mernik_2005}
M.~Mernik, J.~Heering, and A.~M. Sloane, ``When and how to develop
  domain-specific languages,'' \emph{ACM Computing Surveys}, vol.~37, no.~4,
  pp. 316--344, 2005.

\bibitem{Czarnecki_2000}
K.~Czarnecki, U.~Eisenecker, R.~Gl\"{u}ck, D.~Vandevoorde, and T.~Veldhuizen,
  ``Generative programming and active libraries,'' in \emph{Lecture Notes in
  Computer Science (LNCS)}, vol. 1766, 2000, pp. 25--39.

\bibitem{Vandevoorde_2017}
D.~Vandevoorde, N.~Josuttis, and D.~Gregor, \emph{{C++} Templates: The Complete
  Guide}, 2nd~ed.\hskip 1em plus 0.5em minus 0.4em\relax Addison-Wesley, 2017.

\bibitem{Watt_2004}
D.~A. Watt, \emph{Programming Language Design Concepts}.\hskip 1em plus 0.5em
  minus 0.4em\relax Wiley, 2004.

\bibitem{Stock_2012}
K.~Stock, L.-N. Pouchet, and P.~Sadayappan, ``Using machine learning to improve
  automatic vectorization,'' \emph{ACM Transactions on Architecture and Code
  Optimization}, vol.~8, no.~4, 2012.

\bibitem{Cebrian_2020}
J.~M. Cebrian, L.~Natvig, and M.~Jahre, ``Scalability analysis of {AVX-512}
  extensions,'' \emph{The Journal of Supercomputing}, vol.~76, no.~3, pp.
  2082--2097, 2020.

\bibitem{ensmallen2021}
R.~R. Curtin, M.~Edel, R.~G. Prabhu, S.~Basak, Z.~Lou, and C.~Sanderson, ``The
  ensmallen library for flexible numerical optimization,'' \emph{Journal of
  Machine Learning Research}, vol.~22, no. 166, 2021.

\bibitem{mlpack2023}
R.~R. Curtin, M.~Edel, O.~Shrit \emph{et~al.}, ``mlpack 4: a fast, header-only
  {C++} machine learning library,'' \emph{Journal of Open Source Software},
  vol.~8, no.~82, 2023.

\bibitem{rcpparma}
D.~Eddelbuettel and C.~Sanderson, ``{RcppArmadillo}: Accelerating {R} with
  high-performance {C++} linear algebra,'' \emph{Computational Statistics and
  Data Analysis}, vol.~71, pp. 1054--1063, 2014.

\bibitem{Laurent_2004}
A.~{St.~Laurent}, \emph{Understanding Open Source and Free Software
  Licensing}.\hskip 1em plus 0.5em minus 0.4em\relax O'Reilly Media, 2004.

\bibitem{Li_2025}
X.~Li, Y.~Zhang, C.~Osborne \emph{et~al.}, ``Systematic literature review of
  commercial participation in open source software,'' \emph{ACM Transactions on
  Software Engineering and Methodology}, vol.~34, no.~2, 2025.

\bibitem{Henry_2019}
G.~Henry, P.~T.~P. Tang, and A.~Heinecke, ``Leveraging the {bfloat16}
  artificial intelligence datatype for higher-precision computations,'' in
  \emph{IEEE Symposium on Computer Arithmetic (ARITH)}, 2019, pp. 69--76.

\bibitem{Curtin_2025}
R.~R. Curtin, M.~Edel, and C.~Sanderson, ``{Bandicoot}: A templated {C++}
  library for {GPU} linear algebra,'' \emph{arXiv:2508.11385}, 2025.

\bibitem{Hicks_2024}
M.~T. Hicks, J.~Humphries, and J.~Slater, ``{ChatGPT} is bullshit,''
  \emph{Ethics and Information Technology}, vol.~26, no.~2, 2024.

\end{thebibliography}

\end{document}